\documentclass[11pt,a4paper]{article}
\usepackage{graphics}
\usepackage{epsfig}
\usepackage{fancyhdr}
\usepackage{amssymb}
\usepackage{amsmath}

\usepackage[T1]{fontenc}
\usepackage[utf8]{inputenc}
\usepackage{authblk}

\title{Rapidity dependence of particle densities in pp and AA collisions}

\author[1,2]{Irais Bautista\thanks{irais@fpaxp1.usc.es}}
\author[2]{Carlos Pajares\thanks{pajares@fpaxp1.usc.es}}
\author[1,3]{Jos\'e Guilherme Milhano\thanks{guilherme.milhano@ist.utl.pt}}
\author[1]{Jorge Dias de Deus\thanks{jorge.dias.de.deus@ist.utl.pt}}

\affil[1]{CENTRA, Instituto Superior T\'ecnico, Universidade T\'ecnica de Lisboa, \\ Av. Rovisco Pais, P-1049-001 Lisboa, Portugal}
\affil[2]{IGFAE and Departamento de F\'isica de Part\'iculas, Univ. of Santiago de Compostela, 15782, Santiago de Compostela, Spain}
\affil[3]{Physics Department, Theory Unit, CERN, CH-1211 Gen\`eve 23, Switzerland}

\begin{document}

\maketitle

\begin{abstract}
We use multiple scattering and energy conservation arguments to describe $dn/d\eta|_{N_{A}N_{A}}$ as a function of $dn/d\eta|_{pp}$ in the framework of string percolation. We discuss the pseudo-rapidity $\eta$ and beam rapidity $Y$ dependence of particle densities. We present our results for pp, Au-Au, and Pb-Pb collisions at RHIC and LHC.

\end{abstract}

\section{Introduction}

As nuclei are made up of nucleons it is natural to look at nucleus-nucleus (A-A) collisions as resulting from the superposition of nucleon-nucleon (p-p) collisions, in the spirit of Glauber model approach and generalizations of it.
In the single scattering limit the average number of participating nucleons per nucleus, $N_{A}$ behave incoherently and 
\begin{equation}
\frac{dn}{dy}|_{N_A N_A}= \frac{dn}{dy}|_{pp} N_A
\end{equation}
Eq (1) corresponds to the wounded nucleon model \cite{A.Bialas:1976}\cite{N.Armesto.Pajares:200} \cite{J.Dias.R.Ugoccioni:2000}. This model is expected to dominate at very low energy. In general, data do not agree with (1). 

At higher energy one has to take into account multiple scattering and one finds 
 \begin{equation}
 \frac{dn}{dy}|_{N_A N_A}= \frac{dn}{dy}|_{pp} (N^{{1+\alpha(s)}}-N_A).
 \end{equation}
where $N_A^{1+\alpha(s)}$ is the estimated total number of nucleon-nucleon collisions and single scattering was subtracted  \cite{Bautista:2012tn}.

It should be noticed that energy momentum  conservation constrains the combinatorial factors of the Glauber calculus at low energy. The problem is that the energy momentum of $N_A$ valence strings has to be shared by $N_{A}^{4/3}$ (mostly) sea strings. There are proposals to cure this problem, for instance, by reduction of the height of the rapidity plateau for sea strings \cite{Bautista:2011su}. In the same spirit, but reducing the effective number of sea strings rather than reducing the sea plateau, we write (see \cite{Bautista:2012tn}).
 \begin{equation}
 N_{A}^{4/3}\rightarrow N_A^{(1+\alpha(s))}
 \end{equation}
with 
\begin{equation}
\alpha(s)=\frac{1}{3}(1-\frac{1}{1+ln(\sqrt{s/s_0}+1)}),
\end{equation}
such that for $\sqrt{s}<<\sqrt{s_0}$, $\alpha(\sqrt{s})\rightarrow 0$, we are back to the wounded nucleon model, and for $\sqrt{s}>>\sqrt{s_{0}}$, $\alpha(\sqrt{s})\rightarrow \frac{1}{3}$, and we have fully developed Glauber calculus. The need to take multiple scattering contribution was experimentally shown at RHIC \cite{B.B.Back:2000}.

Here as in \cite{N.Armesto.Pajares:200} \cite{J.Dias.R.Ugoccioni:2000} our framework is the Dual Parton Model with parton saturation, and we work with Schwinger strings, with fusion and percolation \cite{N. Armesto:1996}.

In pp and Au-Au collisions or in general $N_A N_A$ collisions the interactions occur with the formation of longitudinal strings in rapidity. The particle density $dn/dy$ is expected to be proportional to the average number of strings (twice the number of elementary collisions) $N^{s}_{N_A} $ (see \cite{Bautista:2012tn}),
\begin{equation}
\frac{dn}{dy}|_{N_{A}N_{A}}\sim \bar{N}^s_{N_{A}}.
\end{equation}

The string percolation model describes the multi particle production in terms of color strings stretched between the partons of the projectile and the target.
 In the impact parameter plane due to the confinement, the color of strings is confined to small area in transverse space $S_{1}=\pi r_{0}^{2}$ with $r_{0} \sim .2-.3$ fm, these strings decay into new ones by $q\bar{q}-\bar{qq}$ pair production and subsequently hadronize to produce the observed hadrons. In the impact parameter plane the strings appear as disks and as energy-density increases the discs overlap, fuse and percolate, leading to the reduction of the overall color \cite{M.A.Braun:2000}\cite{C.Pajares:2007}\cite{C.Pajares:2005}.  A cluster of $n$ strings behaves as a single string with energy momentum corresponding to the sum of the individual ones. An essential quantity is the color reduction factor
\begin{equation}
F(\eta_{N_A}^{t})=\sqrt{ \frac{1-e^{-\eta_{N_A}^{t}}}{\eta_{N_A}^{t}}},
\end{equation}

 where $\eta^{t}_{N_{A}}$ is the string density in the impact parameter plane for $N_A N_A$ collisions given as (see \cite{Bautista:2012tn}):
\begin{equation}
\eta_{N_A}^{t} \equiv \frac{\pi r_{0}^{2}}{S_{N_A}}\bar{N}_{N_A}^{s},
\end{equation} 
$S_{N_A}$ is the area of the impact parameter projected overlap region of the interaction covered by $N_A$ nucleons from nucleus A. Note that
$N_{N_A}^{s}=N_{p}^{s}N_{A}^{1+\alpha}$
and instead of (5) we have now 
\begin{equation}
\frac{dn}{dy}|_{N_A N_A}\sim F(\eta_{N_A}^{t})\bar{N}_{N_A}^{s}
\end{equation}
The color reduction factor $F(\eta_{N_A}^{t})$ is a tool to slow down the increase of $dn/dy$ with energy and number of participating nucleons. Note that in (1) nucleons interact incoherently and $S_{N_A}$ is in fact $S_{p}$, while in (2), due to coherence, $S_{N_A}$ is the overall area of interaction. For details see \cite{Bautista:2012tn}.
We finally have at $\eta=0$, 
\begin{equation}
\frac{1}{N_A}\frac{dn}{d y}|_{N_A N_A}= \kappa \frac{dn}{dy}|_{pp}[1+\frac{F(\eta_{N_A}^{t})}{F(\eta_{p}^{t})}(N_{A}^{\alpha(\sqrt{s})}-1)],
\end{equation}
with $\kappa$ being a normalization factor, 
\begin{equation}
\eta_{N_A}^{t}=\eta_{p}^{t} N_{A}^{\alpha}(\frac{A}{N_{A}^{2/3}}),
\end{equation}
and $F(\eta_{N_A,p}^{t}) \rightarrow \frac{1}{\sqrt{\eta_{N_A,p}^{t}}}$ and $\alpha(\sqrt{s}) \rightarrow \frac{1}{3}$, where $N_{p}^{s}$ is the number of proton strings. At low energy $N_{p}^{s}$ is around 2 growing with energy as $e^{2 \lambda Y}$ (faster than $\frac{dn}{dy}|_{pp}$) so that we can approximately write 
\begin{equation}
N_{p}^{s}=2+4(\frac{r_0}{R_{p}})^{2} e^{2\lambda Y},
\end{equation}

We now generalize the results obtained in ref \cite{Bautista:2012tn}. 

Based on the good description on data obtained by using the formula (9) for different atomic number and number of participants for different energies at mid rapidity, we now apply the same formalism as used in pp to describe the rapidity evolution as suggested in ref \cite{Bautista:2011su}\cite{Brogueira:2009ut}\cite{DiasdeDeus:2007wb} obtaining a general formula for pseudo-rapidity dependence of AA collisions: 

\begin{equation}
\label{eq:dndetaAA}
	\frac{1}{N_A} \frac{dn^{N_AN_A}_{\rm ch}}{d\eta}\bigg|_{\eta} =  \kappa' J \, F(\eta^{t}_{p})\, N_p^s \frac{ \bigg(1+\frac{F(\eta^{t}_{N_{A}})}{F(\eta^{t}_{p})}(N^{\alpha(\sqrt{s})}_{A}-1)\bigg)\ }{\exp{(\frac{\eta-(1-\alpha)Y}{\delta})}+1}
\end{equation}
where $J$ is the usual Jacobean $J=\frac{cosh \eta}{\sqrt{k_{1}+sinh^{2} \eta}}$ and $\kappa'= \frac{\kappa }{J(\eta=0)}( \exp{(\frac{-(1-\alpha)Y}{\delta})}+1)$.

We now apply the formula to describe the charge multiplicity in p-p collisions for different energies in pseudo rapidity,
From our general formula (12) by using  $N_{A}=1$ and $A=1$, to consider p-p collisions the expression is reduced to 
\begin{equation}
 \frac{dn^{pp}_{\rm ch}}{d\eta}\bigg|_{\eta} =  \kappa' \, F(\eta^{t}_{p})\, N_p^s \frac{1}{\exp{(\frac{\eta-(1-\alpha)Y}{\delta})}+1}
\end{equation}

\section{Comparison with experimental data (RHIC, LHC)}

In figure (1) It is shown the comparison of the formula (13) applyied to different energies at different pseudo-rapidities with data from different experiments and energies, showing a good agreement in the evolution in pseudo rapidity and an increase in the plateau region as increasing with energy.
\begin{figure}[h]
\begin{center}
      \resizebox{140mm}{!}{\includegraphics{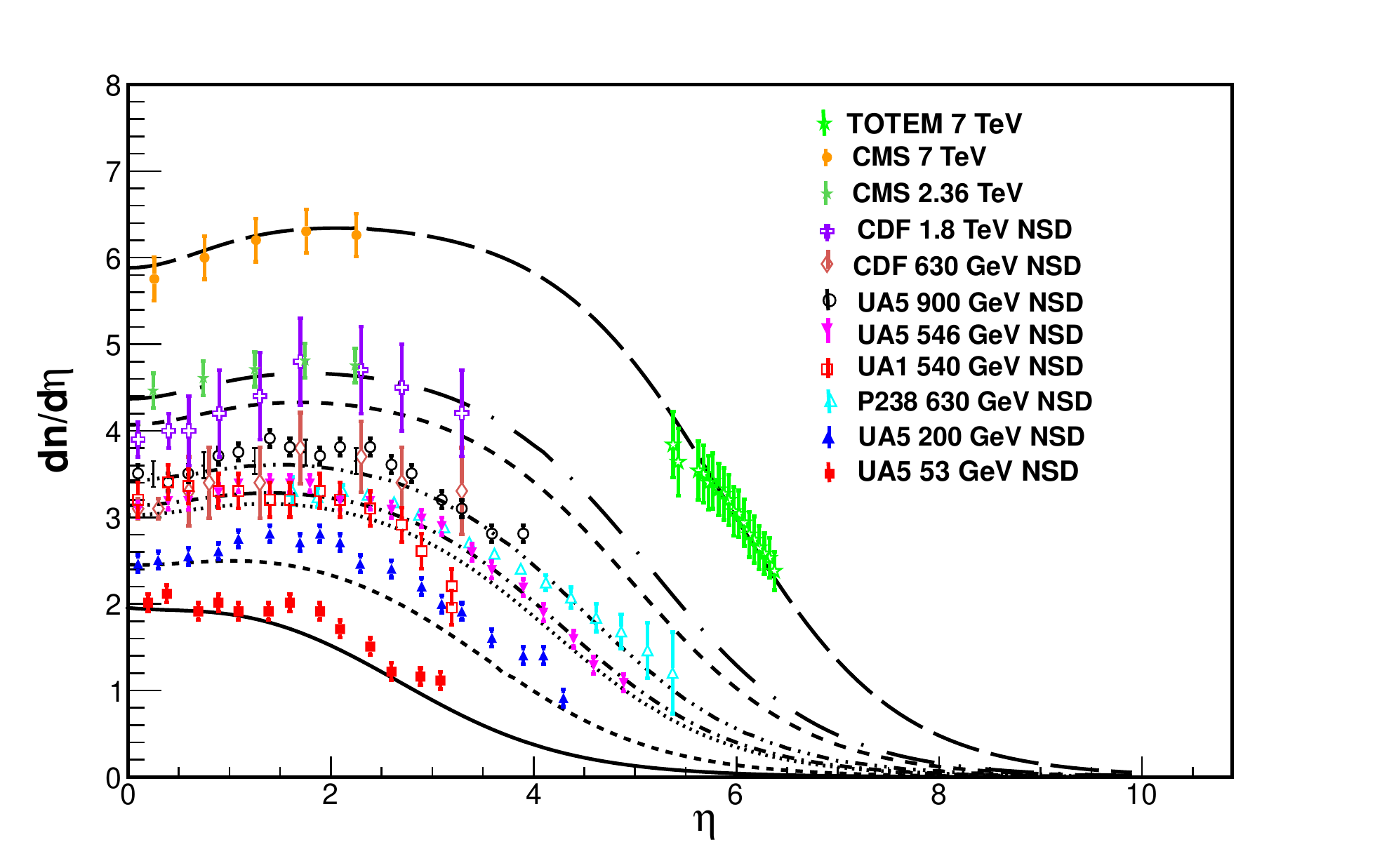}}
       \caption[Comparison of the results from the evolution of the $dn_{ch}/d\eta$ with the
     pseudorapidity ]{Comparison of the results from the evolution of the $dn_{ch}/d\eta$ with dependence in
     pseudorapidity from equation (13)  for p-p collisions at different energies (lines),  data is taken from ref. \cite{pp1data} \cite{pp2data} \cite{Antchev:2012ux}.}
\end{center}
\end{figure}

In figure (2), (3), (4) and (5)  it is shown the comparison between our results from formula (12) for Cu-Cu, Au-Au and Pb-Pb collisions  at different energies, in agreement with data.

In figure (6) We show some predictions for  3.2, 3.9 and 5.5 TeV energies at centrality $0-5\%$, for Pb-Pb collisions.

In the above computations we have used the following values of the parameters:
 $\kappa  = 0.63 \pm 0.01$, $\lambda = 0.201 \pm 0.003$, and $\sqrt{s_{0}} = 245 \pm 29 $ GeV, the same as obtained in \cite{Bautista:2012tn}, to describe the particle density $\frac{dn}{d\eta}|_{N_AN_A}$ in the same power law as  $\frac{dn}{d\eta}|_{pp}$. We had made here an extension to these descriptions to add the pseudo rapidity evolution with the same aim as in ref \cite{Bautista:2011su}.
 
 The new parameters values $\alpha \simeq 0.34 $, $\delta \simeq 0.84 $, $k_{1}=1.2$  had been set to adjust the equation (13) with data [13][14[15].  These results can be  extended to describe proton-nucleus collisions.   
\newpage
\section{Conclusions}
We have discussed in a general way the physics of particle densities in pp and AA collisions. Our model gives a non-linear dependence of $\frac{1}{N_A}dn/d\eta|_{AA}$ on $dn/d\eta|_{pp}$. Particle densities, as a function of $\eta$ and $Y$
give a good description of Pb-Pb 
data at LHC in a wide region in $\eta$. The same is observed for pp in a wide range of rapidity $Y$. 

Notice that recent data from TOTEM experiment measurements in the the charged particle pseudorapidity density $dN_{ch}/d\eta$ in pp collisions at $\sqrt{s} = 7$ TeV for $5.3 < |\eta| < 6.4$  have been compared to several MC generators and none of them has been found to fully describe the measurement, but our model is able to reproduce it.

\begin{figure}[h]
\begin{center}
    \begin{tabular}{cc}
     \resizebox{100mm}{!}{\includegraphics{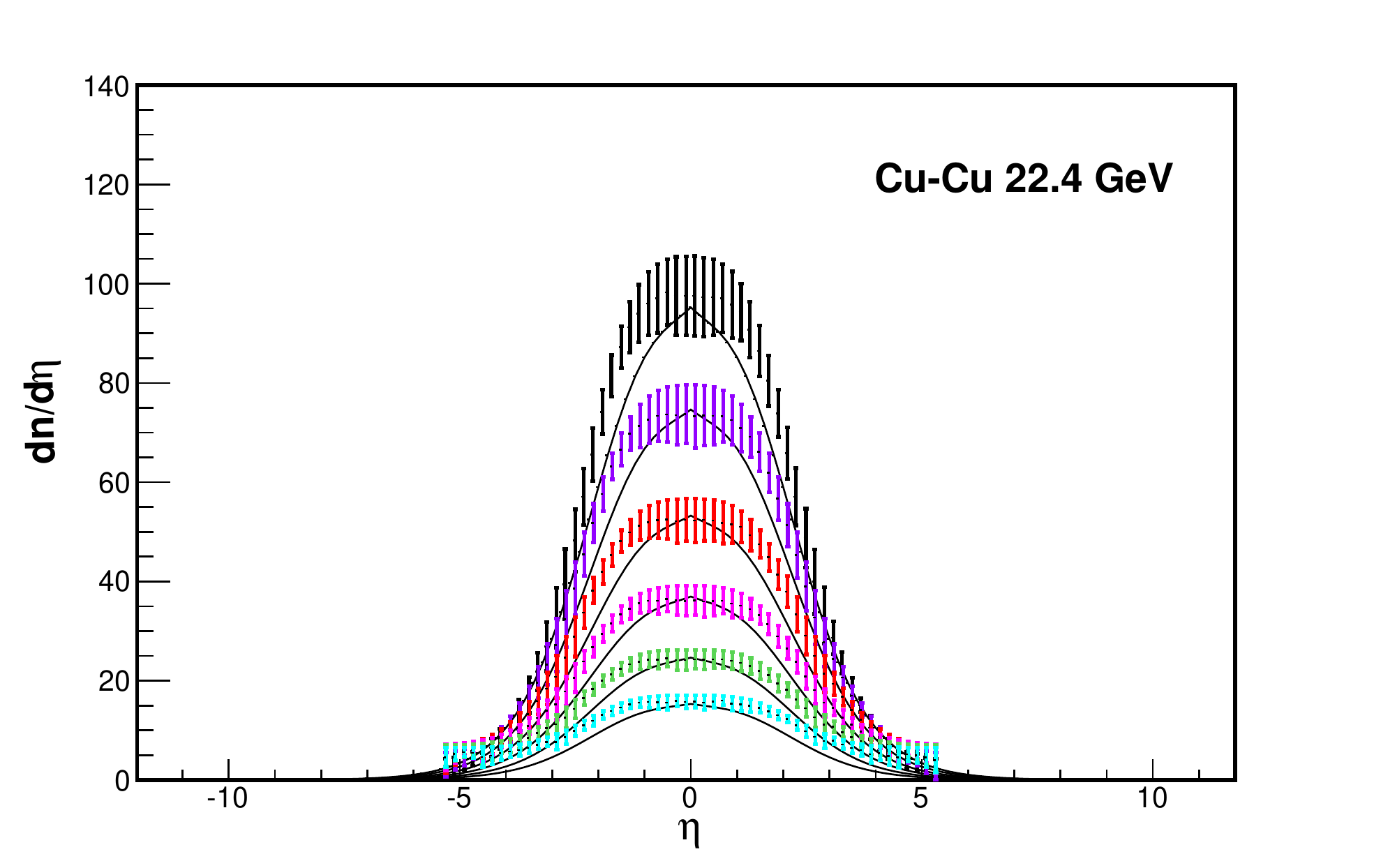}}\\
     \resizebox{100mm}{!}{\includegraphics{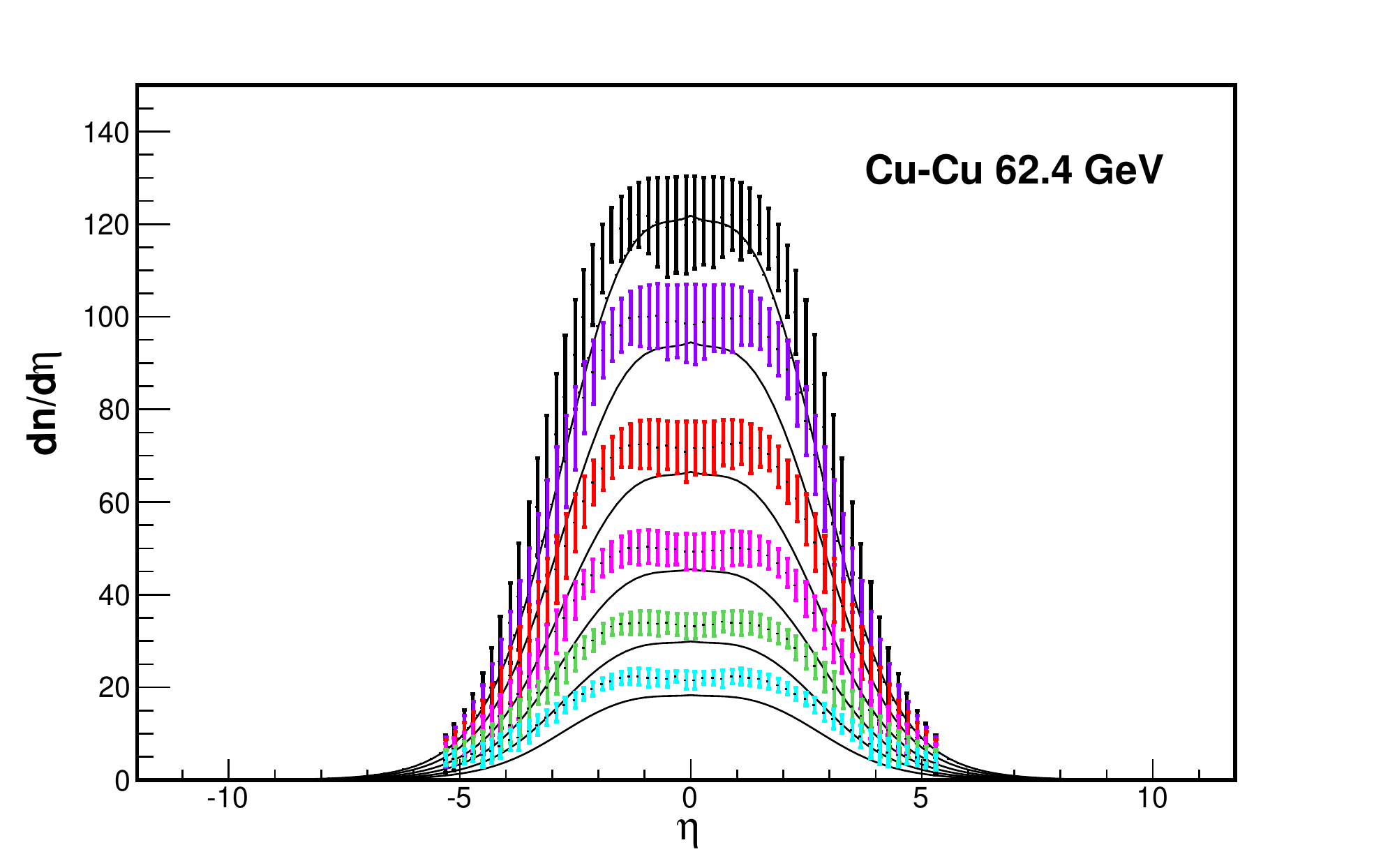}}\\
     \resizebox{100mm}{!}{\includegraphics{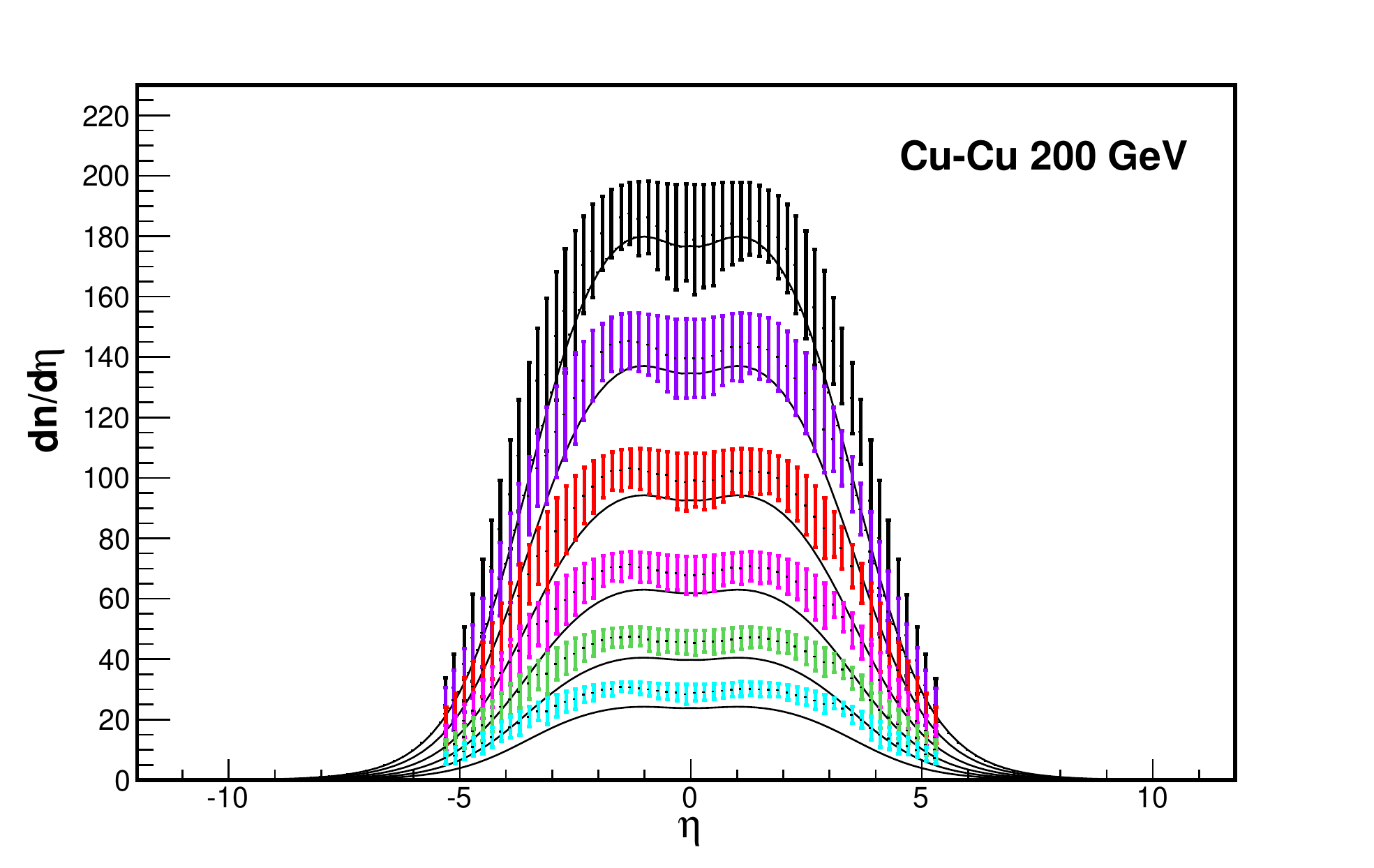}}\\
    \end{tabular}
    \caption[Comparison of the results from the evolution of the $dn_{ch}/d\eta$ with the
     pseudorapidity ]{Comparison of the results from the evolution of the $dn_{ch}/d\eta$ with
     pseudorapidity from equation (12) for Cu-Cu collisions at 22.4 GeV, 62.4 GeV, and 200 GeV energies, data is taken from ref. \cite{Alver:2007aa}., Error bars in color blue, green, pink, red, purple and black are used for the corresponding centralities $45-55\%$, $35-45\%$, $25-35\%$, $15-25\%$, $6-15\%$, $0-6\%$ respectively, lines in black show our results. }

\end{center}     

\end{figure}

\begin{figure}[h]
\begin{center}
    \begin{tabular}{cc}
    
     \resizebox{100mm}{!}{\includegraphics{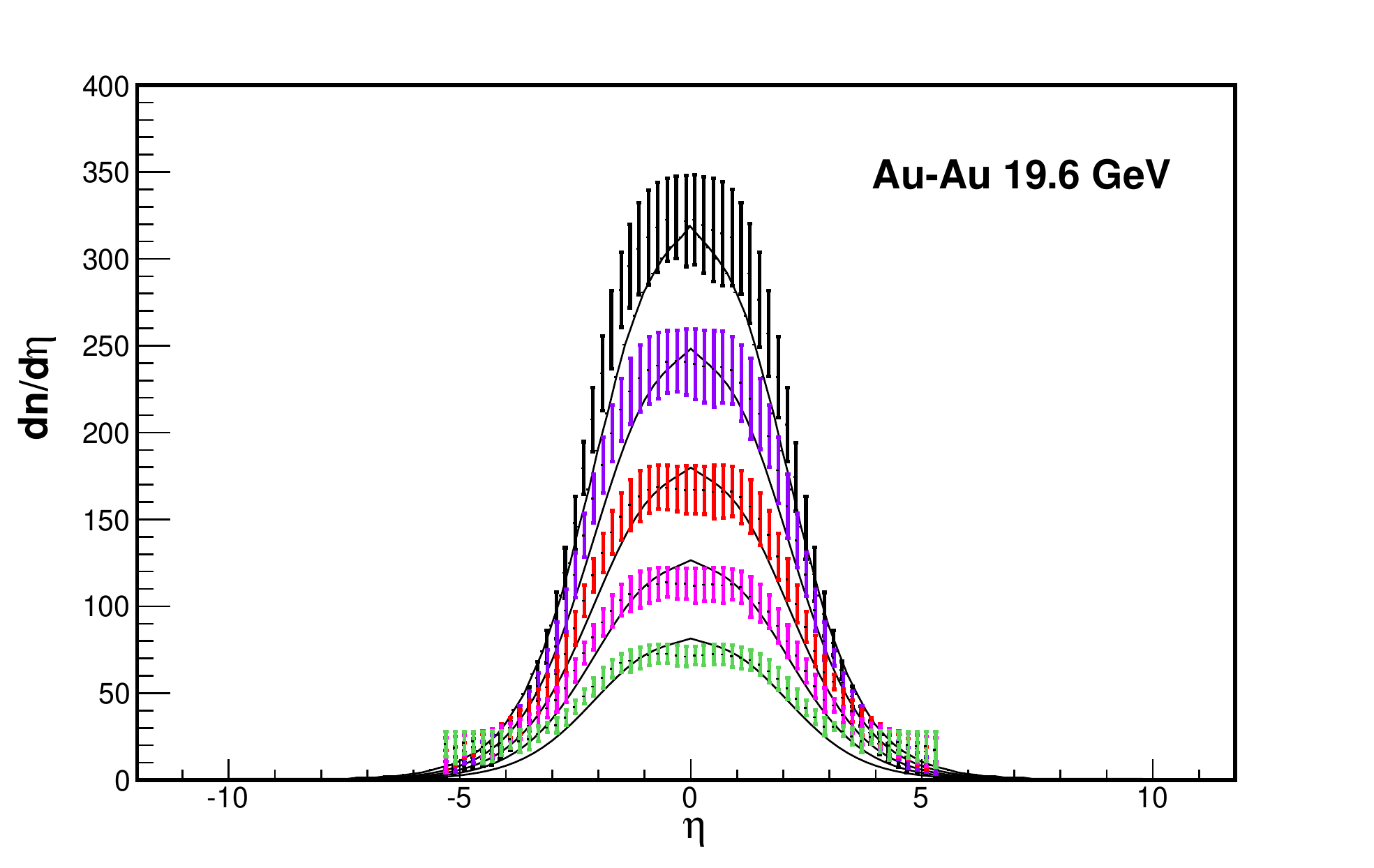}}\\
     \resizebox{100mm}{!}{\includegraphics{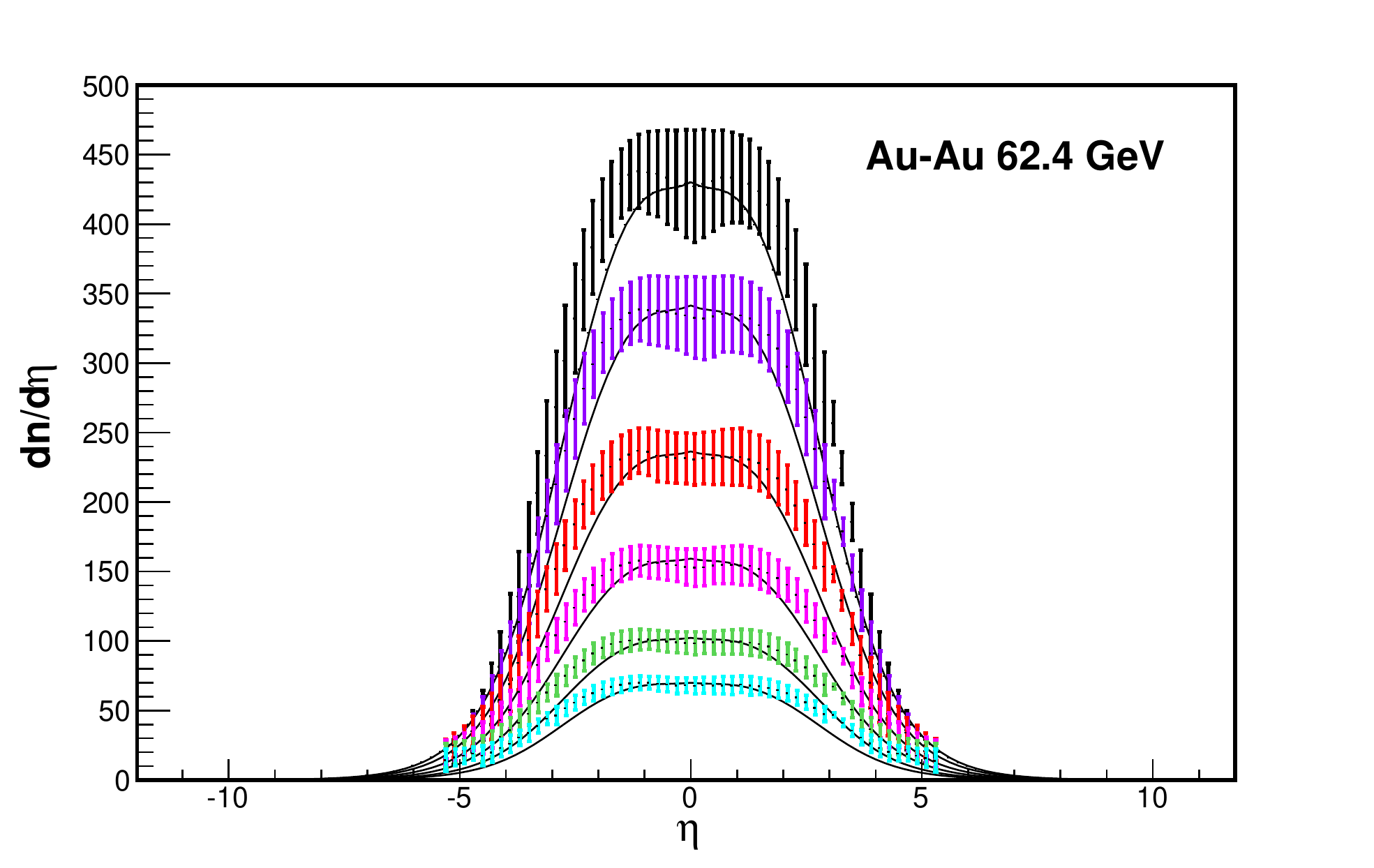}}\\
     \resizebox{100mm}{!}{\includegraphics{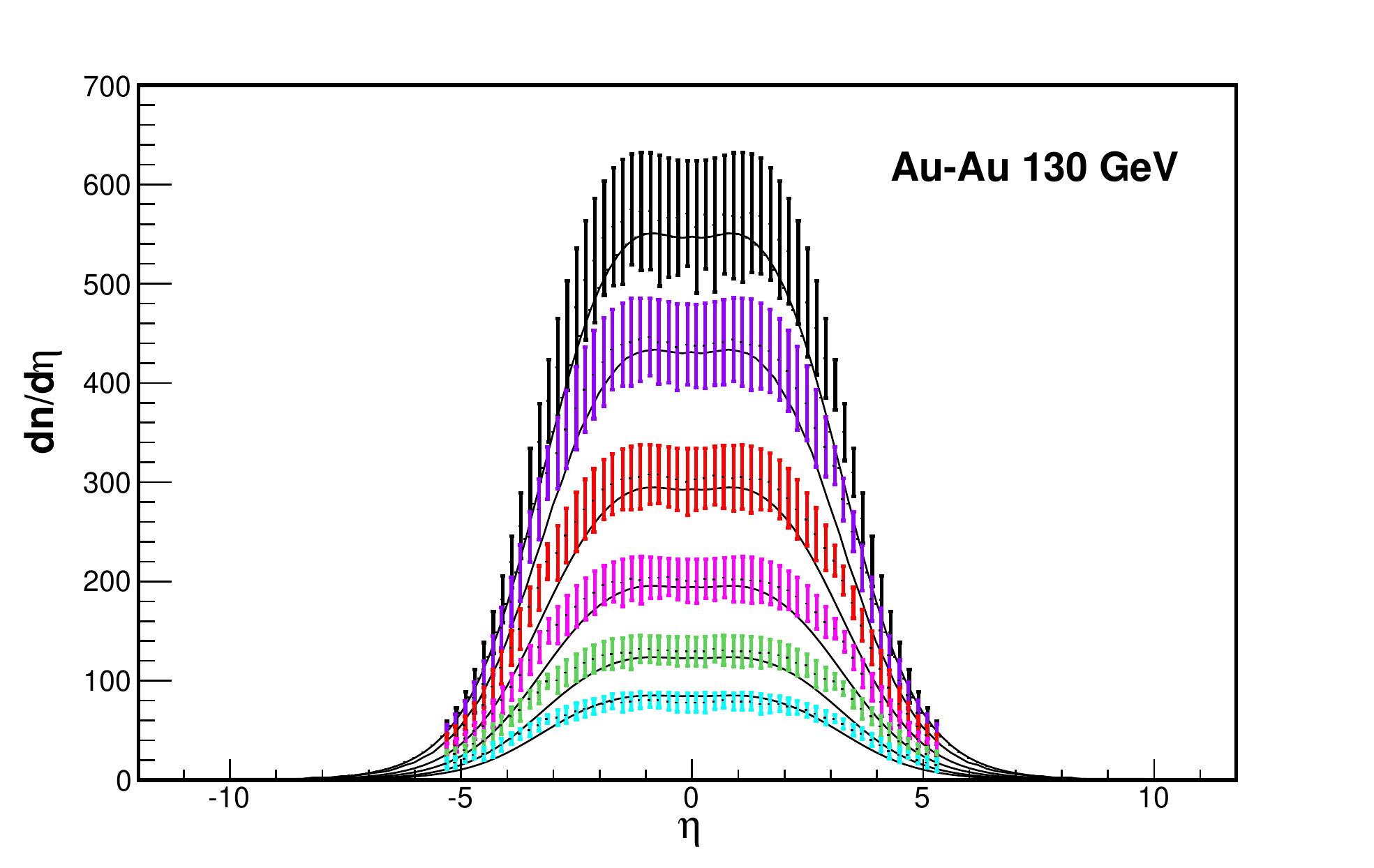}}\\

    \end{tabular}
    \caption[Comparison of the results from the evolution of the $dn_{ch}/d\eta$ with the
     pseudo-rapidity ]{Comparison of the results from the evolution of the $dn_{ch}/d\eta$ with
     pseudorapidity from equation (12) for Au-Au collisions at 19.6 GeV, 62.4 GeV and
      130 GeV energies, data is taken from \cite{data2AuAu}.
       Error bars in color blue, green, pink, red, purple and black are used for the corresponding
        centralities $45-55\%$, $35-45\%$, $25-35\%$, $15-25\%$, $6-15\%$, $0-6\%$ respectively, lines in black are the model results. }
\end{center}     

\end{figure}
\begin{figure}
\begin{center}
     \resizebox{100mm}{!}{\includegraphics{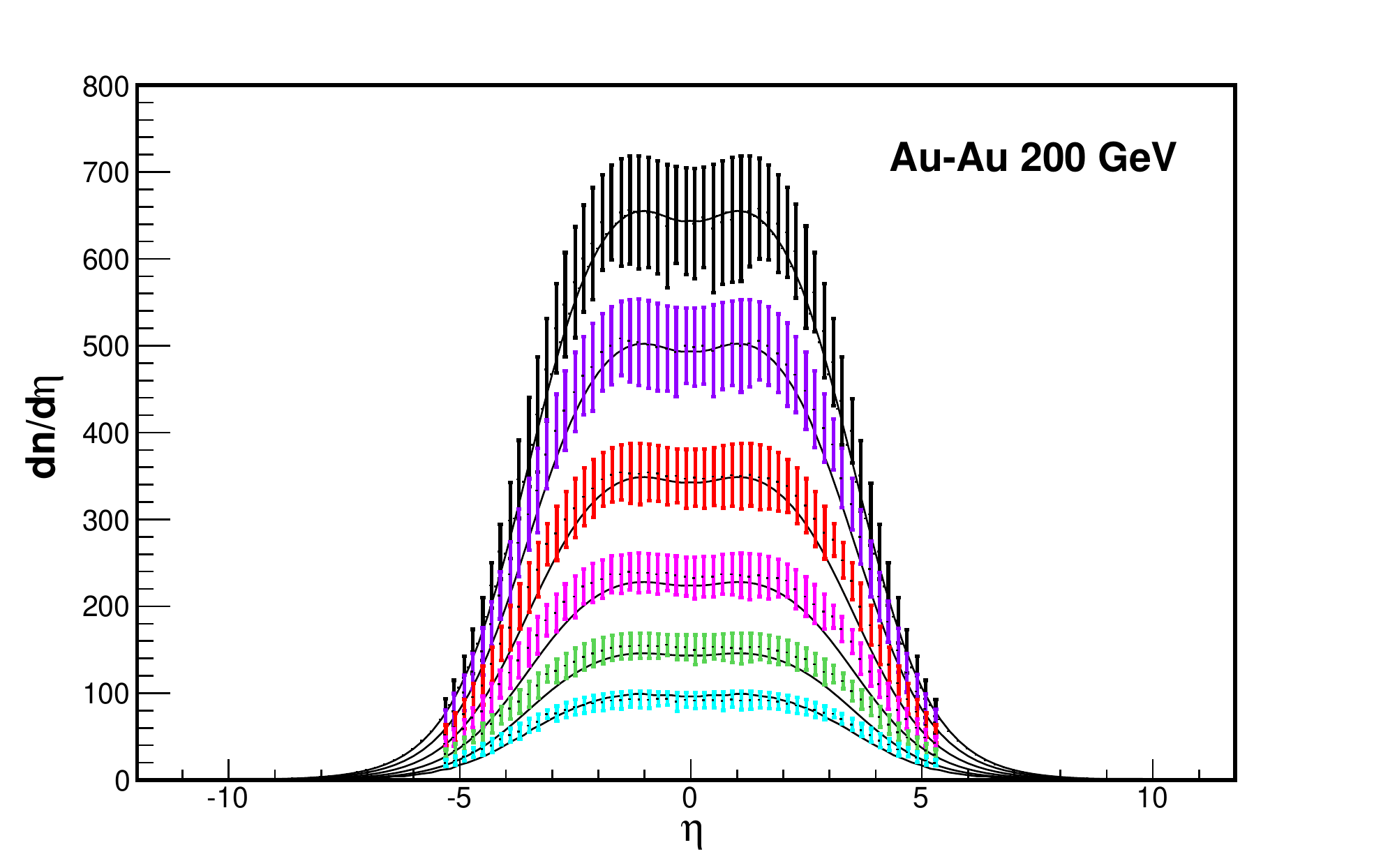}}\\
    \caption[Comparison of the results from the evolution of the $dn_{ch}/d\eta$ with the
     pseudorapidity ]{Comparison of the results from the evolution of the $dn_{ch}/d\eta$ in
     pseudorapidity from equation (12) for Au-Au collisions at  200 GeV, data is taken from \cite{data2AuAu}. Error bars in color blue, green, pink, red, purple and black are used for the  corresponding centralities $45-55\%$, $35-45\%$, $25-35\%$, $15-25\%$, $6-15\%$, $0-6\%$ respectively, lines in black are the model results. }
\end{center}     
\end{figure}

\begin{figure}
\begin{center}
      \resizebox{120mm}{!}{\includegraphics{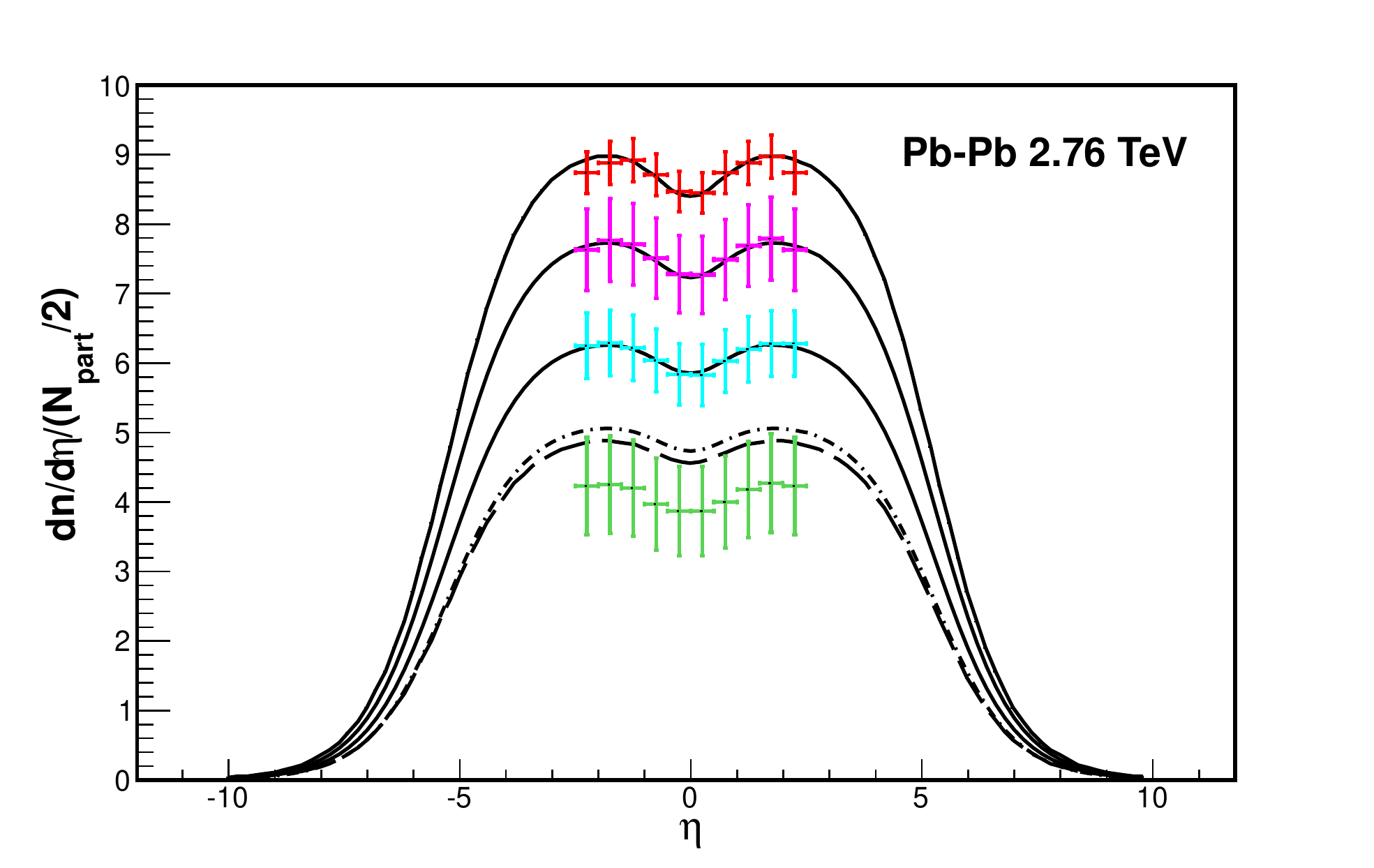}}\\
     \caption[Comparison of the results from the evolution of the $(dn_{ch}/d\eta)/(N_{part}/2)$ with the
     pseudorapidity ]{Comparison of the results from the evolution of the $(dn_{ch}/d\eta)/(N_{part}/2)$ with the
     pseudorapidity from equation (12) for Pb-Pb collisions at 2.76 TeV, data is taken from \cite{Chatrchyan:2011pb}. Error bars in color green,  blue, pink and red are used for the corresponding centralities  $85-95\%$, $50-55\%$, $0-90\%$ and $0-5\%$ respectively, lines in black are the corresponding results from the model to the respective centrality, for the smaller centrality we use the number of participants corresponding to the 85-95$\%$ showed in dot and dashed line and in dashed line is the minimum number of participants equal 2.  
      }
    \label{test4}
\end{center}
\end{figure}

\begin{figure}
\begin{center}
      \resizebox{120mm}{!}{\includegraphics{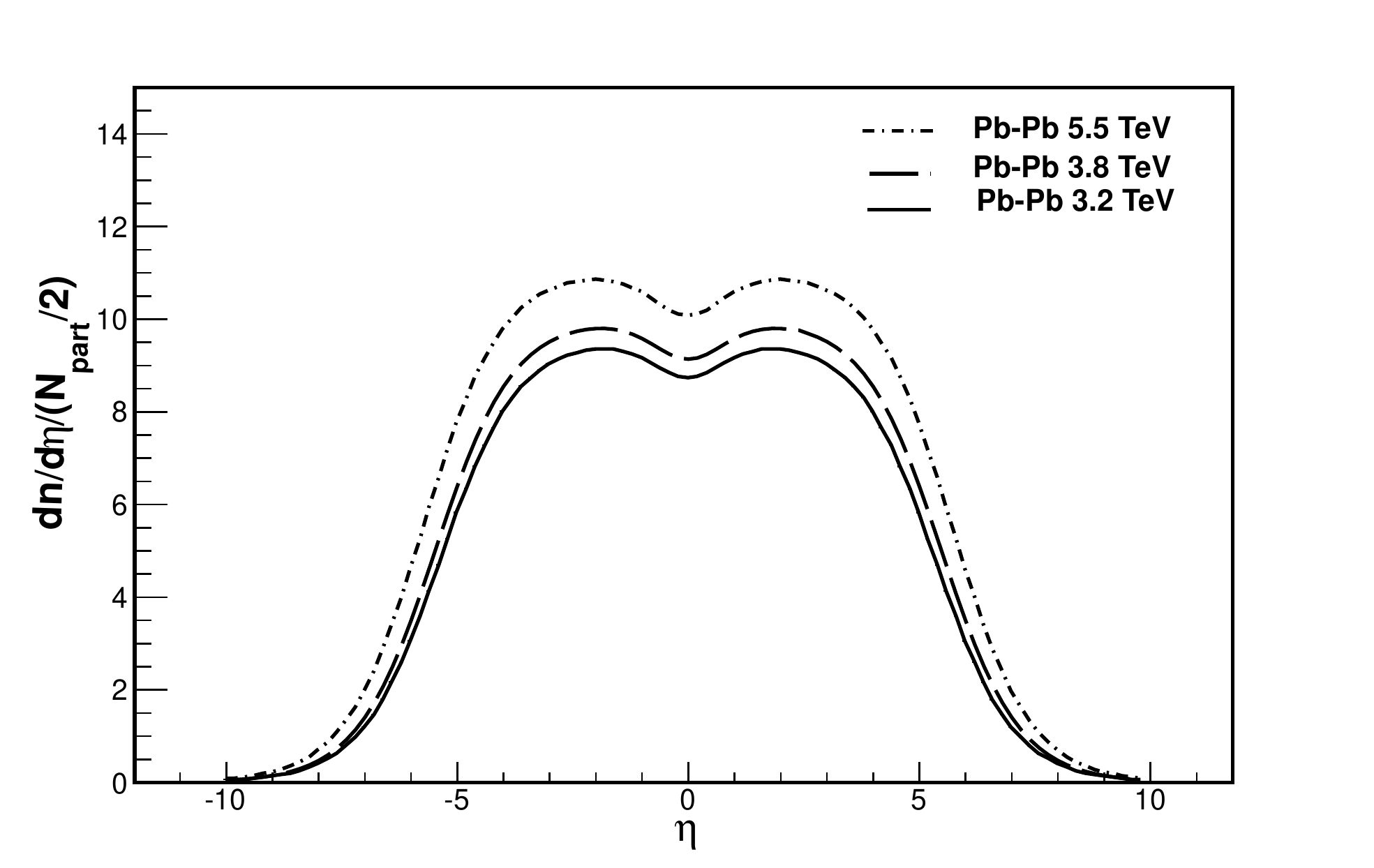}}\\
     \caption[Precitions on the evolution of the $\frac{dn_{ch}/d\eta}{N_{part}/2}$ with the
     pseudo-rapidity ]{Predictions on the evolution of the $(dn_{ch}/d\eta)/(N_{part}/2)$ with
     pseudorapidity from equation (12) for Pb-Pb collisions at  3.2, 3.9 and 5.5 TeV energies at  $0-5\%$ centrality. }
    \label{test4}
\end{center}
\end{figure}

\newpage

\section*{Acknowledgments}

IB is supported by  the grant SFRH/BD/51370/2011 from Funda\c c\~ao para a Ci\^encia e a Tecnologia (Portugal).
JDD and JGM acknowledge the support of Funda\c c\~ao para a Ci\^encia e a Tecnologia (Portugal) under project CERN/FP/116379/2010. IB and CP were partly supported by the project FPA2008-01177  and FPA2011-22776 of MICINN, the Spanish Consolider Ingenio 2010 program CPAN and Conselleria de Educacion Xunta de Galicia.

\end{document}